\documentclass[12pt]{article}
\title{Extremal black holes and nilpotent orbits}
\usepackage{mathrsfs}
\usepackage{amsmath}

\global\arraycolsep=1pt
\usepackage[colorlinks=true,backref=true,linkcolor=black,anchorcolor=black,citecolor=black,filecolor   =black,menucolor=black,pagecolor=black,urlcolor=black]{hyperref}
\usepackage{bbm}
\usepackage{dsfont}
\usepackage{amssymb}
\usepackage{textcomp}
\usepackage{wasysym}
\DeclareMathAlphabet{\mathpzc}{OT1}{pzc}{m}{it}

\newcommand{\Scal}[1]{\Bigl ({#1} \Bigr )}
\newcommand{\scal}[1]{\bigl ({#1} \bigr )}
\def\be{\begin{equation}}
\def\ee{\end{equation}}

\def\DJo{$\;$\kern-.4em \hbox{D\kern-.8em\raise.15ex\hbox{--}\kern.35em okovi\'c}}

\newcommand{\eprint}[1]{{\texttt{[#1]}}}
\newcommand{\eprintN}[1]{{\texttt{#1 [hep-th]}}}

\def\ie{{\it i.e.}\ }

\def\V{{\mathcal{V}}}
\def\w{{\scriptstyle W}}

\def\G{{\mathfrak{G}}}
\def\H{{\mathfrak{H}}}

\def\M{{\mathcal{M}}}

\def\asym{{\scriptscriptstyle 0}}

\newcommand{\ord}[1]{{\scriptscriptstyle (#1)}}

\def\n{{\mathpzc{n}}}

\def\zero{{\mathpzc{0}}}
\def\un{{\mathpzc{1}}}
\def\deux{{\mathpzc{2}}}
\def\trois{{\mathpzc{3}}}

\def\invo{{\APLstar}}

\DeclareMathOperator{\ad}{ad}

\newcommand{\trace}{\hbox {Tr}~}

\def\e{\boldsymbol{e}}

\def\h{\boldsymbol{h}}

\def\G{\mathfrak{G}}
\def\h{\mathfrak{h}}
\def\H{\mathfrak{H}}

\def\w{{\scriptstyle W}}

\def\gl{\mathfrak{gl}}

\def\so{\mathfrak{so}}

\def\e{\mathfrak{e}}

\def\N{\mathcal{N}}

\begin{document}
\allowdisplaybreaks[1]
\renewcommand{\thefootnote}{\fnsymbol{footnote}}
\begin{titlepage}
\begin{flushright}
\
\vskip -2.5cm
{\small AEI-2009-099}\\
\vskip 1cm
\end{flushright}
\begin{center}
{\Large \bf
Extremal black holes and nilpotent orbits}
\\
\lineskip .75em
\vskip 3em
\normalsize
{\large  Guillaume Bossard\footnote{email address: bossard@aei.mpg.de}}\\
\vskip 1 em
$^{\ast}${\it AEI, Max-Planck-Institut f\"{u}r Gravitationsphysik\\
Am M\"{u}hlenberg 1, D-14476 Potsdam, Germany}
\\

\vskip 1 em
\end{center}
\begin{abstract}
The stationary solutions of a large variety of (super)gravity theories can be described within a non-linear sigma model $\G/\H^*$ coupled to Euclidean gravity in three-dimensions, for which $\G$ is a simple group and $\H^*$ a non-compact real form of its maximal compact subgroup. The absence of naked singularities in four dimensions requires the $\G$ Noether charge in 3D to satisfy a characteristic equation that determines it in function of the mass, the NUT charge and the electro-magnetic charges of the solution.  It follows that  the Noether charge associated to extremal black holes must lie in a certain Lagrangian submanifold of a nilpotent orbit of $\G$. Constructing  a suitable parameterisation of this  Lagrangian, we are able to determine the so-called `fake superpotential' that governs the radial dependency of the scalar fields.
\end{abstract}

\end{titlepage}
\renewcommand{\thefootnote}{\arabic{footnote}}
\setcounter{footnote}{0}


In trying to extend our  quantum gravity understanding of black holes, extremality is often a key simplifying assumption. Firstly, it eliminates Hawking radiation and ensures that the solution is semi-classically stable. Secondly, it guarantees that the near-horizon 
solution is entirely determined by the conserved charges measurable at spatial infinity,
and therefore insensitive (away from lines of marginal stability) to variations to the asymptotic value of the scalar fields at infinity. This attractor behaviour, first discovered for supersymmetric (BPS) black holes \cite{Ferrara:1995ih,Ferrara:1996um}, holds for all extremal solutions \cite{Ferrara:1997tw,Sen:2005wa, Goldstein:2005hq}, and is arguably responsible for the validity of certain weakly coupled description of non-BPS black hole micro-states \cite{Dabholkar:2006tb} in string theory. 

Considering supergravity theories, black holes are solutions of the Einstein equations coupled to abelian vector fields and scalar fields parameterising a Riemannian space $\M_4$. Assuming spherical symmetry and extremality, the supergravity equations of motion become equivalent to light-like geodesic motion on a pseudo-Riemannian space $\M_3^*$, with the affine parameter $\tau$ identified as the inverse radial distance $\tau=1/r$. Restricting ourselves to static solutions, or equivalently considering asymptotically Minkowski solutions, the equations of motion reduce furthermore to the motion of a fiducial particle on $\mathbb{R}^*_+ \times \M_4$ subject to a negative definite $V$ depending on the electromagnetic charges.  There are two main approaches to determine such solutions that we combined in \cite{Bossard:2009we}.

The first one consists in reducing the equations of motion in first order gradient flow equations through the determination of a `fake superpotential' $W$ satisfying
\be
V=  - e^{2U} \left( W^2+2 g^{ij}\partial_i W\partial_{j}W \right) \label{vbhsuper}
\ee
where $g_{ij}$ is the Riemannian metric on $\M_4$ and $U$ is the scale function defining the extremal static spherically symmetric metric ansatz
\be
\label{4dmetric}
ds^2 = -e^{2U} dt^2+ e^{-2U} \scal{ dr^2 + r^2 (d\theta^2 +\sin^2\theta \, d\phi^2)} 
\ee
Whenever the solution is BPS, \ie that it preserves some supersymmetry generators, the first order gradient flow equations correspond to the vanishing of the supersymmetry transformation of the fermion  fields.   This is most familiar in the framework of $D=4, \N=2$ supergravity, for which the BPS superpotential is determined in function of the central charge $Z$ as 
\be  W_{\rm \scriptscriptstyle BPS} = |Z_{p,q}(\phi^i)| \ee
This permitted to determine the most general BPS solutions explicitly \cite{Denef:2000nb,Bates:2003vx}. However, the solution to (\ref{vbhsuper}) is not unique, and the determinations of inequivalent `fake superpotentials' $W$ were obtained in \cite{Ceresole:2007wx,Andrianopoli:2007gt,
LopesCardoso:2007ky,Perz:2008kh,Hotta:2007wz,Gimon:2007mh,Bellucci:2008sv,Gimon:2009gk}.

On the other-hand, when $\M_4$ is a symmetric space $\G_4 / \H_4$, such that its isometry group $\G_4$ defines a symmetry of the theory by acting faithfully on the electromagnetic field strength, the pseudo-Riemannian manifold $\M_3^*$ on which is defined the light-like geodesic motion is itself a symmetric space $\G/\H^*$ \cite{Breitenlohner:1987dg}, with simple isometry group $\G$ and a non-compact maximal subgroup $\H^*$ which is a non-compact real form of the maximal compact subgroup of $\G$.  The geodesic motion on $\M_3^*$ is then integrable, and in fact 
all geodesics on $\M_3^*$ can be obtained by exponentiating
a generator $ - P_\asym \tau  \in \mathfrak{g} \ominus \mathfrak{h}^*$, where
$P_\asym$ determines the momentum along the trajectory. $P_\asym$ is conjugate to the Noether
charge $Q$ via the coset representative $\V$ in $\G/ \H^*$,
\be 
P \equiv -  \Scal{\V^{-1} \, Ê\dot{\V}   }  \big|_{\mathfrak{g} \ominus \mathfrak{h}^*} = \V^{-1} Q \V 
\ee
where the dot denote the derivative with respect to $\tau$.  Extremal solutions correspond to  special geodesics which reach the boundary $U=-\infty$ in infinite proper time \cite{Breitenlohner:1987dg}. It is necessary but not sufficient that the geodesic be light-like.
For BPS black holes, it was observed in \cite{Gunaydin:2005mx} that the Noether
charge must satisfy $[\ad(Q)]^5=0$, \ie belong to a nilpotent orbit
of degree 5. More recently, the supersymmetry and extremality conditions on the Noether charge 
for symmetric supergravity models were re-analyzed in \cite{Bossard:2009at}.
It was shown in all cases where $\G$ is simple that extremality requires 
to $[Q_{|\bf R}]^3=0$, where $\bf R$ denotes the ``fundamental representation" of $\G$:
for example  the spinor representation if $\G$ is an orthogonal group $SO(2+m,2+n)$ or $SO^*(2m+4)$.
The only exception is for $\G = E_{8(8)}$ or $E_{8(-24)}$, where
the condition becomes $[Q_{|{\bf 3875}}]^5=0$, with $\bf 3875$ being the $3875$-dimensional irreducible representation appearing in the symmetric tensor product of two adjoints. 

More precisely, any generic extremal spherically symmetric black hole (\ie with a non-zero horizon area) is characterized by a nilpotent Noether charge $Q$ which lies inside the grade-two component 
$ \mathfrak{l}_4^\ord{2}$ of $\mathfrak{g}$ with respect to the 5-grading (more appropriately, even 9-grading) which arises in the reduction from 4 to 3 dimensions:
\be \mathfrak{g} \cong {\bf 1}^\ord{-4}Ê\oplus \mathfrak{l}_4^\ord{-2}Ê\oplus \scal{Ê\gl_1 \oplus \mathfrak{g}_4}^\ord{0} \oplus  \mathfrak{l}_4^\ord{2} \oplus  {\bf 1}^\ord{4} 
\label{DimRedGrade} 
\ee
The nilpotent orbit $\mathcal{O}_{\G}$ of $Q\in \mathfrak{g}$ under $\G$ is  
characterized by the isotropy subgroup  \footnote{for extremal black holes, this isotropy subgroup coincides with the isotropy subgroup of the electromagnetic charges in the four-dimensional duality group $\G_4$ computed in 
\cite{Levi,Ferrara:1997uz,Bellucci:2006xz}.} of $Q$ 
in $\G$. On the other hand, the momentum $P_\asym$ is valued in the coset 
$\mathfrak{g} \ominus \mathfrak{h}^*$, and therefore defines a $\H^*$-orbit $\mathcal{O}_{\H^*}$ 
inside $\mathfrak{g} \ominus \h^*$.

Since the coset component of the Maurer--Cartan form 
is conjugate to the Noether charge via $P = \V^{-1} Q \V$, 
it defines a representative ${\bf e} \equiv P$ of the corresponding nilpotent orbit inside the coset component $\mathfrak{g} \ominus \mathfrak{h}^*$, and therefore defines a $\H^*$-orbit inside this coset. A general fact
about nilpotent elements is that one can always find another nilpotent element ${\bf f}$ and
a semi-simple generator ${\bf h}$ such that the triplet
$({\bf e},\, {\bf f},\, {\bf h})$ defines an $\mathfrak{sl}_2$ subalgebra
of $\mathfrak{g}$, \ie 
\be
\label{triple}
[{\bf e},{\bf f}]={\bf h} \qquad [{\bf h},{\bf e}]=2{\bf e} \qquad [{\bf h},{\bf f}]=- 2{\bf f} 
\ee
The eigenspaces of ${\bf h}$ furnish a graded decomposition of $\mathfrak{g}$ which uniquely 
characterizes the complex nilpotent $\G_{\mathbb{C}}$ orbit \cite{Collingwood}. Extremal solutions are such that the $\H^*$-orbit of $P$ is characterized by a graded decomposition of $\mathfrak{h}^*$ of the 
 same form as (\ref{DimRedGrade}) \cite{Bossard:2009my}, 
\be \mathfrak{h}^* \cong \mathfrak{h}^\ord{-4} \oplus \mathfrak{h}^\ord{-2} \oplus \gl_1 \oplus \mathfrak{h}^\ord{0} \oplus  \mathfrak{h}^\ord{2} \oplus  \mathfrak{h}^\ord{4} 
\ee
In the case of maximally supersymmetric supergravity \cite{CremmerJulia}, for static solutions 
(\ie with zero NUT charge) the semi-simple  element $\bf h$ associated to 
the nilpotent element $P$ can be computed in terms of the central charges $Z_{ij}$ alone, and more generally, in terms of the central and matter charges which we write collectively $Z_I$.\footnote{Here $Z_I$ are the scalar field dependent linear combinations of the electromagnetic charges, 
transforming in a complex representation of $\H_4$ and such that $V_{\rm \scriptscriptstyle BH}=Z_I Z^I$.} 
Decomposing  $P\in \mathfrak{g}\ominus \mathfrak{h}^*$ with respect to  the Ehlers  $U(1)$ and the four-dimensional R-symmetry group $\H_4$ \cite{Bossard:2009we},
 \be 
P = - \dot{U}\,  {\bf H} \, + \, e^U Z_I\,  {\bf L}^I  \, - \,  {e_i}^j \dot{\phi}^i \, {\bf G}_j   
\quad \in \quad  \mathbb{C} \oplus \mathfrak{l}_4 \oplus \scal{Ê\mathfrak{g}_4 \ominus \mathfrak{h}_4} 
\ee
where ${e_i}^j$ is a vielbein for the metric $g_{ij}$, 
one may recast the middle equation in \eqref{triple}
into a system of first order differential 
equations of the form
\be \dot{U} =  - e^{U} W\,   \qquadÊ 
g_{ij}Ê\,  \dot{\phi}^j =  - e^{U} W_i 
\label{FirstOrderSys} 
\ee
where $W$ and $W_i$ depend on the moduli $\phi^i$ and electromagnetic charges $Q_I$ through
the charges $Z_I$ only; moreover, we proved that \cite{Bossard:2009we}
\be
 W_i = \partial_{\phi^i} W 
 \ee
Thus, extremal solutions attached to the given nilpotent orbit 
satisfy  a gradient flow under the fake superpotential $W$. 
In particular, it follows from the nilpotency of $P$ that 
\be \trace P^2 = 0 = e^{2U} \Scal{ÊW^2 - Z_I Z^I + 2 g^{ij} W_i W_j} = 0\ , 
\ee
and therefore that \eqref{vbhsuper} is obeyed.

Applying this strategy to $\N=8$ supergravity with $\G=E_{8(8)}$, we are able to determine the fake superpotentials for both BPS and non-BPS extremal black holes, and express them in terms of the $SU(8)$ invariant combinations of the central charges. In this case $P_\asym$ transforms  as a Majorana--Weyl spinor under $Spin^*(16)$. It can be  conveniently parameterised using a fermionic oscillator basis \cite{Bossard:2009at},
\begin{multline} 
\hspace{-4mm} 
|P_\asym \rangle = \Scal{Ê\w + Z_{ij} a^i a^j + \Sigma_{ijkl} a^i a^j a^k a^l + \frac{1}{6!} \varepsilon_{ijklmnpq} \scal{ÊZ^{pq} \, a^i \cdots  a^n +  \frac{1}{56} \bar \w  \, a^i \cdots  a^q } } | 0 \rangle  \\*
= ( 1 + \invo ) \Scal{ÊÊ\w + Z_{ij} a^i a^j + 
\frac{1}{2}Ê\Sigma_{ijkl} a^i a^j a^k a^l } | 0 \rangle 
\end{multline} 
where $\invo$ is the anti-involution defining the chiral Majorana--Weyl representation 
of $Spin^*(16)$,  $\w = M + i N$ where $M$ is the mass and $N$ the NUT charge, 
$Z_{ij}$ are the supersymmetric central charges  and $\Sigma_{ijkl}$ are
the  ``scalar charges". 

There are two $E_{8(8)}$ orbits associated to the nilpotency condition  $[Q_{|{\bf 3875}}]^5=0$, whose union is dense in the space of solutions of this equation \cite{Bossard:2009at,E8strat}. They both lie in a single $E_8(\mathbb{C})$ orbit, associated to the same five graded decomposition,
\be 
\e_{8(8)} \cong   {\bf 1}^{\ord{-4}} 
\oplus {\bf 56}^{\ord{-2}} \oplus \scal{Ê\gl_1 \oplus 
\e_{7(7)}}^\ord{0} \oplus {\bf 56}^\ord{2} 
\oplus {\bf 1}^\ord{4} \label{fiveTwo}
\ee
A representative ${\bf E}$ of such a nilpotent orbit is a generic element of the grade two component ${\bf 56}^\ord{2}$. There are two classes of such elements which are distinguished by their isotropy subgroup inside $E_{7(7)}$, respectively $E_{6(2)}$ and $E_{6(6)}$ \cite{Levi,Ferrara:1997uz}.  The $\gl_1$ generator ${\bf h}$ of  $\so^*(16)$  which defines a corresponding graded decomposition of $\so^*(16)$ and its Majorana--Weyl representation ${\bf 128}_+$ are such that a representative of the orbit lies in the component of grade two of ${\bf 128}_+$
\be 
{\bf h} \, | P_\asym \rangle = 2 | P_\asym \rangle
\label{GradeTwo}
\ee
The two orbits of $Spin^*(16)$ associated to generic extremal black holes (\ie black holes with a non-vanishing horizon area) correspond to the 1/8 BPS and the non-BPS extremal black holes, respectively.

In order for the solution to be supersymmetric, the corresponding Noether 
charge state must satisfy the `Dirac equation' \cite{Bossard:2009at} 
\be\label{Dirac1} 
\scal{Ê\epsilon_\alpha^i a_i + 
\varepsilon_{\alpha\beta} \epsilon^\beta_i a^i } | P_\asym \rangle = 0 
\ee
where $a^i$ and $ a_j$ (for $i,j,\dots = 1,...,8$) are the fermionic 
oscillators from which the spinor representations of $Spin^*(16)$ are built. In the case of an 1/8 BPS solutions, this equation permits to determine the generator ${\bf h}$ from the value of the central charge as 
\be 
\label{omom}
{\bf h}_{\frac{1}{8}} \equiv Êe^{i \varphi} 
R_i{}^k R_j{}^l \, Ê\omega^\zero_{kl} a^i a^j -  Êe^{-i \varphi} 
R^i{}_k R^j{}_l \, Ê\omega^{\zero\, {kl}} a_i a_j
\ee
for a central charge decomposing on a basis of real orthonormal antisymmetric tensor $\omega^\n_{ij}$ as
\be
Z_{ij} = \frac{1}{2}Êe^{i \varphi} R_i{}^k R_j{}^l \Scal{Ê\rho_\zero \omega^\zero_{ij} + \rho_\un \, \omega^\un_{ij} +  \rho_\deux \, \omega^\deux_{ij} +   \rho_\trois \, \omega^\trois_{ij} }
\ee
such that $\rho_\zero \ge \rho_\un \ge \rho_\deux \ge \rho_\trois$. This determines the BPS `fake superpotential' to be 
\be W_{\scriptscriptstyle \rm BPS} = \rho_\zero(Z_{ij}) \ee
in agreement with  \cite{Andrianopoli:2007gt}.   

In the non-BPS case, one use the fact that the generator $\mathfrak{h}$ also define the nilpotent elements associated to 1/2 BPS solutions by its grade four component, to use the `Dirac equation' (\ref{Dirac1}) to determine ${\bf h}$ in this case as well. As a result, 
\be  Ê{\bf h }Ê=  \frac{1}{2 \cos (2 \alpha) } \Scal{Ê e^{-i \alpha} \tilde{R}_i{}^k  \tilde{R}_j{}^l  \Omega_{kl}  a^i a^j - e^{i \alpha} \tilde{R}^i{}_k  \tilde{R}^j{}_l  \Omega^{kl} a_i a_j }   + i  \tan(2\alpha)  \scal{Êa^i a_i - 4 } \ee
for a central charge decomposing as  
\be
 Z_{ij} =  e^{-i\alpha}Ê\tilde{R}_i{}^k \tilde{R}_j{}^l \biggl( Ê\, \frac{1}{2} \Scal{Êe^{2 i\alpha} + i \sin(Ê2\alpha)  } \varrho \, \Omega_{kl} +   \Xi_{kl} \, \biggr)   \label{ZNB}  \ee
such that $\Omega_{ij}$ is a symplectic form of overall phase factor $e^{i\frac{\pi}{4}}$ and $\Xi_{ij}$ satisfies
\be 
\Xi_{ij} =   \Omega_{ik} \Omega_{jl} \Xi^{kl} \  \qquad 
\Omega^{ij}  \Xi_{ij} = 0 
 \label{real27} 
\ee
The `fake superpotential' is thus defined in the non-BPS case by
\be  W=2\varrho(Z_{ij}) \ee
$\varrho$ being also a particular root of an irreducible sextic  polynomial \cite{Bossard:2009we}.

By truncation, one then obtains the `fake superpotential' for all magic $\N=2$ supergravity
models, and in fact for all supergravity theories with $\N \ge 2$ with a symmetric moduli space.



\begin{thebibliography}{9}


\bibitem{Ferrara:1995ih}
  S.~Ferrara, R.~Kallosh and A.~Strominger,
  ``$\N=2$ extremal black holes,''
  Phys.\ Rev.\  D {\bf 52} (1995) 5412
  \eprint{hep-th/9508072}.

\bibitem{Ferrara:1996um}
  S.~Ferrara and R.~Kallosh,
  ``Universality of supersymmetric attractors,''
  Phys.\ Rev.\  D {\bf 54} (1996) 1525
  \eprint{hep-th/9603090}.

\bibitem{Ferrara:1997tw}
  S.~Ferrara, G.~W.~Gibbons and R.~Kallosh,
  ``Black holes and critical points in moduli space,''
  Nucl.\ Phys.\  B {\bf 500} (1997) 75
  \eprint{hep-th/9702103}.

\bibitem{Sen:2005wa}
  A.~Sen,
  ``Black hole entropy function and the attractor mechanism in higher
  derivative gravity,''
  JHEP {\bf 0509} (2005) 038
  \eprint{hep-th/0506177}.
  
  
\bibitem{Goldstein:2005hq}
  K.~Goldstein, N.~Iizuka, R.~P.~Jena and S.~P.~Trivedi,
  ``Non-supersymmetric attractors,''
  Phys.\ Rev.\  D {\bf 72} (2005) 124021
  \eprint{hep-th/0507096}.
  

\bibitem{Dabholkar:2006tb}
  A.~Dabholkar, A.~Sen and S.~P.~Trivedi,
  ``Black hole microstates and attractor without supersymmetry,''
  JHEP {\bf 0701} (2007) 096
  \eprint{hep-th/0611143}.
  
  \bibitem{Bossard:2009we}
  G.~Bossard, Y.~Michel and B.~Pioline,
  ``Extremal black holes, nilpotent orbits and the true fake superpotential,''
  \eprintN{0908.1742}.


  

\bibitem{Denef:2000nb}
  F.~Denef,
  ``Supergravity flows and D-brane stability,''
  JHEP {\bf 0008} (2000) 050
  [arXiv:hep-th/0005049].

\bibitem{Bates:2003vx}
  B.~Bates and F.~Denef,
  ``Exact solutions for supersymmetric stationary black hole composites,''
  arXiv:hep-th/0304094.


\bibitem{Ceresole:2007wx}
  A.~Ceresole and G.~Dall'Agata,
  ``Flow equations for non-BPS extremal black holes,''
  JHEP {\bf 0703}, 110 (2007)
  \eprint{hep-th/0702088}.

\bibitem{Andrianopoli:2007gt}
  L.~Andrianopoli, R.~D'Auria, E.~Orazi and M.~Trigiante,
  ``First order description of black holes in moduli space,''
  JHEP {\bf 0711} (2007) 032
  \eprintN{0706.0712}.

\bibitem{LopesCardoso:2007ky}
  G.~Lopes Cardoso, A.~Ceresole, G.~Dall'Agata, J.~M.~Oberreuter and J.~Perz,
  ``First-order flow equations for extremal black holes in very special
  geometry,''
  JHEP {\bf 0710} (2007) 063
  \eprintN{0706.3373}.


\bibitem{Perz:2008kh}
  J.~Perz, P.~Smyth, T.~Van Riet and B.~Vercnocke,
  ``First-order flow equations for extremal and non-extremal black holes,''
  JHEP {\bf 0903} (2009) 150
  \eprintN{0810.1528}.


\bibitem{Hotta:2007wz}
  K.~Hotta and T.~Kubota,
  ``Exact solutions and the attractor mechanism in non-BPS black holes,''
  Prog.\ Theor.\ Phys.\  {\bf 118} (2007) 969
  \eprintN{0707.4554}.


\bibitem{Gimon:2007mh}
  E.~G.~Gimon, F.~Larsen and J.~Simon,
  ``Black holes in supergravity: the non-BPS branch,''
  JHEP {\bf 0801} (2008) 040
  \eprintN{0710.4967}.

\bibitem{Bellucci:2008sv}
  S.~Bellucci, S.~Ferrara, A.~Marrani and A.~Yeranyan,
  ``$stu$ black holes unveiled,''
  \eprintN{0807.3503}.
  
\bibitem{Gimon:2009gk}
  E.~G.~Gimon, F.~Larsen and J.~Simon,
  ``Constituent Model of Extremal non-BPS Black Holes,''
  JHEP {\bf 0907} (2009) 052
  \eprintN{0903.0719}.


\bibitem{Breitenlohner:1987dg}
  P.~Breitenlohner, D.~Maison and G.~W.~Gibbons,
  ``Four-dimensional black holes from Kaluza--Klein theories,''
  Commun.\ Math.\ Phys.\  {\bf 120} (1988) 295.




\bibitem{Gunaydin:2005mx}
  M.~G\"unaydin, A.~Neitzke, B.~Pioline and A.~Waldron,
  ``BPS black holes, quantum attractor flows and automorphic forms,''
  Phys.\ Rev.\  D {\bf 73} (2006) 084019
  \eprint{hep-th/0512296}.


\bibitem{Bossard:2009at}
  G.~Bossard, H.~Nicolai and K.~S.~Stelle,
  ``Universal BPS structure of stationary supergravity solutions,''
  \eprintN{0902.4438}.

  \bibitem{Levi}
 D.~\v{Z}.~\DJo, 
 ``Classification of nilpotent elements in simple exceptional real Lie algebras of inner type and description of their centralizers,''
 J.\ of Algebra {\bf 112} (1988) 503. 
  

  \bibitem{Ferrara:1997uz}
  S.~Ferrara and M.~G\"unaydin,
  ``Orbits of exceptional groups, duality and BPS states in string theory,''
  Int.\ J.\ Mod.\ Phys.\  A {\bf 13} (1998) 2075
  \eprint{hep-th/9708025}.

\bibitem{Bellucci:2006xz}
  S.~Bellucci, S.~Ferrara, M.~G\"unaydin and A.~Marrani,
  ``Charge orbits of symmetric special geometries and attractors,''
  Int.\ J.\ Mod.\ Phys.\  A {\bf 21} (2006) 5043
  \eprint{hep-th/0606209}.



\bibitem{Collingwood}
D.~Collingwood and W.~ McGovern,
``Nilpotent orbits in semisimple Lie algebras"
Van Nostrand Reinhold Mathematics Series,  New York, 1993.


\bibitem{Bossard:2009my}
  G.~Bossard and H.~Nicolai,
  ``Multi-black holes from nilpotent Lie algebra orbits,''
  \eprintN{0906.1987}.



 \bibitem{CremmerJulia}
  E.~Cremmer and B.~Julia,
  ``The $SO(8)$ Supergravity,''
  Nucl.\ Phys.\  B {\bf 159} (1979) 141.


\bibitem{E8strat}
D.~\v{Z}.~\DJo,
``The closure diagram for nilpotent orbits of the split real form of $E_8$,''
CEJM {\bf 4} (2003) 573.



\end{thebibliography}
\end{document}